\documentstyle[emulateapj]{article}

\def\lae{\mathrel{<\kern-1.0em\lower0.9ex\hbox{$\sim$}}}
\def\gae{\mathrel{>\kern-1.0em\lower0.9ex\hbox{$\sim$}}}

\slugcomment{Accepted for publication in the Astrophysical Journal, April 10 2000 issue.}
 
\lefthead{C\^ot\'e et al.}
\righthead{Hierarchical Formation of the Galactic Spheroid}
 
\begin{document}
 
\title{Evidence for the Hierarchical Formation of the Galactic Spheroid}
 
\author{Patrick C\^ot\'e\altaffilmark{1}}
\affil{California Institute of Technology, Mail Stop 105-24, Pasadena, CA 91125}

\author{Ronald O. Marzke\altaffilmark{2}}

\affil{Observatories of the Carnegie Institute of Washington,\\
813 Santa Barbara Street, Pasadena, CA 91101}

\smallskip

\author{Michael J. West}

\affil{Department of Astronomy \& Physics, Saint Mary's University, Halifax, 
NS, B3H 3C3, Canada, and \\ Department of Physics \& Astronomy,
University of Hawaii, Hilo, HI 96720}

\and

\author{Dante Minniti}

\affil{Lawrence Livermore National Laboratory,
Livermore, CA 94550, and \\
Departament of Astronomy and Astrophysics, Pontificia
Universidad Cat\'olica de Chile, Casilla 306, Santiago 22, Chile}

\altaffiltext{1}{Sherman M. Fairchild Fellow}
\altaffiltext{2}{Hubble Fellow}

% Notice that each of these authors has alternate affiliations, which
% are identified by the \altaffilmark after each name.  The actual alternate
% affiliation information is typeset in footnotes at the bottom of the
% first page, and the text itself is specified in \altaffiltext commands.
% There is a separate \altaffiltext for each alternate affiliation
% indicated above.
 
% The abstract environment prints out the receipt and acceptance dates
% if they are relevant for the journal style.  For the aasms style, they
% will print out as horizontal rules for the editorial staff to type
% on, so long as the author does not include \received and \accepted
% commands.  This should not be done, since \received and \accepted dates
% are not known to the author.
 
\begin{abstract}
The possibility that the Galactic spheroid was assembled from numerous,
chemically-distinct proto-Galactic fragments is investigated using a Monte-Carlo 
technique designed to simulate the chemical evolution of the Galaxy 
in hierarchical formation scenarios which involve no gas dissipation.
By comparing the observed and simulated metallicity distributions of Galactic 
globular clusters and halo field stars, we estimate the level of fragmentation 
in the collapsing proto-Galaxy.
Although the formation process is highly stochastic, the
simulations frequently show good agreement with the observed metallicity distributions,
provided the luminosity function of proto-Galactic fragments had
the form $dN \propto L^{{\alpha}}dL$ where $\alpha \sim -2$. While this
steep slope is strongly at odds with
the presently observed luminosity function of the Local Group, it is
in close agreement with the predictions of semi-analytic and numerical models
of hierarchical galaxy formation. 
We discuss a number of possible explanations for this discrepancy.
These simulations suggest that the Galactic halo and its globular cluster system
were assembled via the accretion and disruption
of $\sim$ $10^3$ metal-poor, proto-Galactic fragments by the dominant
building block: a proto-bulge whose own metal-rich globular clusters system
has been preferentially eroded by dynamical processes. This formation scenario
may provide a simple explanation for the different shapes of the Galactic globular 
cluster and halo star metallicity distributions.
Based on the similar properties of globular clusters belonging to spiral and giant
elliptical galaxies, we argue that the same process ($e.g.$, hierarchical
growth involving little gas dissipation) is responsible for
the formation of both giant elliptical galaxies and the bulge-halo components
of spiral galaxies.
\end{abstract}
 
% The different journals have different requirements for keywords.  The
% keywords.apj file, found on aas.org in the pubs/aastex-misc directory, 
% contains a list of keywords used with the ApJ and Letters.  These are 
% usually assigned by the editor, but authors may include them in their 
% manuscripts if they wish. 
 
 
% That's it for the front matter.  On to the main body of the paper.
% We'll only put in tutorial remarks at the beginning of each section
% so you can see entire sections together.
 
% In the first two sections, you should notice the use of the LaTeX \cite
% command to identify citations.  The citations are tied to the
% reference list via symbolic KEYs.  We have chosen the first three
% characters of the first author's name plus the last two numeral of the
% year of publication.  The corresponding reference has a \bibitem
% command in the reference list below.
%
% Please see the AASTeX manual for a more complete discussion on how to make
% \cite-\bibitem work for you.   

\section{Introduction}

It is generally believed that galaxy formation begins with the 
collapse of gravitationally unstable, primordial density fluctuations ($e.g.$, 
Gunn \& Gott 1972; Press \& Schechter 1974). In hierarchical formation models, 
such as those involving cold dark matter,
large galaxies then grow at the expense of their smaller counterparts
($e.g.$, White \& Rees 1978; Blumenthal et al. 1984;
Kauffmann, White \& Guiderdoni 1993; Cole et al. 1994). 
Despite the impressive successes of these models, particularly in 
describing the clustering properties of massive galaxies (Baugh et al. 1998),
they have difficulty reproducing the luminosity distribution of nearby galaxies.
For instance, a generic prediction of such hierarchical
models is the existence of a large population of low-mass, dark halos in the local
universe (Kauffmann, White \& Guiderdoni 1993; Klypin et al. 1999). In the specific 
case of the Local Group, the expected number of dark halos having circular velocities
$v_c \lae 30$ km s$^{-1}$ exceeds the observed number of
faint galaxies by an order of magnitude or more (Klypin et al 1999;
Moore et al. 1999).

From an empirical perspective, hierarchical models were anticipated by Searle \& Zinn (1978)
who suggested that the Galactic halo formed via the protracted infall of 
``transient proto-Galactic fragments". As supporting evidence, they
cited the lack of an abundance gradient among the
outer halo globular clusters and the possible spread in age
suggested by the diversity of their horizontal branch morphologies.
The extent to which this scenario differs from that of Eggen, Lynden-Bell \& 
Sandage (1962), who had argued for the rapid collapse of a homogeneous
proto-Galactic gas cloud, is primarily a matter of the formation timescale 
and the properties of the ``transient proto-Galactic fragments" (particularly their 
masses and total numbers) since gravitational and thermal instabilities in 
the collapsing gas cloud must invariably lead to fragmentation into isolated 
star-forming regions (Fall \& Rees 1985; Sandage 1990). Clearly, the key to 
distinguishing between these different scenarios lies
in the duration of the formation process and level of fragmentation 
in the collapsing proto-Galaxy.

In recent years, support has grown for the notion that the halo was, at least
in part, assembled from small, protogalactic fragments as suggested by Searle \& 
Zinn (1978). Such evidence includes the discovery of the tidally-disturbed 
Sagittarius dwarf galaxy (Ibata, Gilmore \& Irwin 1994), numerous reports of 
kinematic substructure
among halo field stars (Preston, Beers \& Schectman 1994; Majewski, Munn \& 
Hawley 1996; Kinman et al. 1996; Chen 1998), improved evidence for age spreads
among the halo field and cluster populations (Laird \& Sneden 1996; Sarajedini, 
Chaboyer \& Demarque 1997; Hesser et al. 1998) and chemical evidence for 
accreted substructure among halo field stars (Carney 1996; King 1997).
Mateo (1996) explored the possibility that the {\it entire} Galactic 
halo was formed through the accretion and disruption of faint
dSph galaxies, assumed to represent the remains of surviving
``Searle-Zinn fragments". Based on a comparison of the stellar populations,
dark matter, variable stars, and globular clusters of the Galactic halo with
those of its present retinue of dSph galaxies, Mateo (1996) concluded that
such a scenario is indeed plausible, provided that the bulk of the accretion 
occured at early times (see also Unavane, Wyse \& Gilmore 1996). By contrast,
van den Bergh (1996) has reviewed the evidence for a rapid, and early,
collapse of a single proto-Galactic cloud in the inner regions of the 
Milky Way.

An associated, but as yet unanswered, question is how the formation of the Milky
Way and other spiral galaxies is related to that of giant elliptical galaxies.
Although it is often assumed that variations in the star formation rate and gas cooling 
efficiency can produce end-products which have markedly different morphologies ($e.g.$, 
Steinmetz \& M\"uller 1995; Haehnelt, Steinmetz \& Rauch 1998), there are some 
intriguing similarities between the Galactic globular clusters system and those of 
giant elliptical galaxies which suggest a closely
related formation history. Most notably, many giant elliptical galaxies contain,
as does the Milky Way, chemically-distinct globular clusters systems ($e.g.$, Whitmore et al. 1995;
Geisler, Lee \& Kim 1996; Forbes, Brodie \& Grillmair 1997; Morgan 1959; Kinman 1959; Zinn 1985).
C\^ot\'e, Marzke \& West (1998) showed that such multi-modal globular clusters metallicity
distributions may be a signature of the formation of giant elliptical galaxies through the
accretion of numerous faint dwarf galaxies and/or proto-galactic fragments.
This scenario is reminiscent of the Searle \& Zinn (1978) model for the
formation of the Galactic halo, and the suggestion by Harris \& Pudritz (1994) 
that supergiant molecular clouds having masses similar to dwarf galaxies were
the sites of globular cluster formation in the early universe.

In this paper, we review and compare the properties of the globular clusters
associated with giant elliptical and spiral galaxies, paying particularly close 
attention to the globular cluster system of the Milky Way.
We then describe a technique to test the possibility that the Galactic halo 
formed via the accretion and disruption of numerous proto-Galactic fragments.

\section{A Comparision of Globular Cluster Systems in Elliptical and Spiral Galaxies}

Renzini (1999) argued that there exists a fundamental connection
between elliptical galaxies and the bulge-halo components of spiral galaxies, 
suggesting that the former can be thought of as spiral bulge-halo systems 
which ``for some reason missed the
opportunity to acquire or maintain a prominent disk". Much evidence now 
supports this view, at least for ellipticals of intermediate-luminosity: 
$e.g$, both classes obey the same Mg$_2$-$\sigma$ relationship
(Jablonka, Martin \& Arimoto 1996),  structural
parameter relations (Kormendy 1985; Bender, Burstein \& Faber 1992) and
$L \propto \sigma^n$ relation (Kormendy \& Illingworth 1983).

In this section, we review briefly the properties of globular clusters systems in spiral
and elliptical galaxies,
paying particular attention to their specific frequencies, 
metallicity distributions and spatial distribution.
In agreement with the conclusions of the above 
studies, we find evidence for a close connection between elliptical
galaxies and the bulge-halo components of spiral galaxies.
{\sl In what follows, we refer to the combined halo-bulge components of spiral 
galaxies as their spheroids}. 

\subsection{Specific Frequencies}

Harris \& van den Bergh (1981) defined the total number of globular clusters per unit 
host galaxy luminosity as the globular cluster specific frequency,
$$S_n = N_{\rm gc}10^{0.4(M_V + 15)}. \eqno{(1)}$$
According to Harris (1991), $S_n \simeq 4-6$ for dwarf and giant 
ellipticals in rich clusters, whereas such galaxies in loose groups
have $S_n \simeq 2-3$. By contrast,
$S_n \simeq 1$ for spiral galaxies. This difference forms the basis 
of the familiar argument that giant elliptical galaxies cannot form 
via spiral-spiral mergers (van den Bergh 1982; cf. Ashman \& Zepf 1992).

The above value of $S_n \simeq 1$ 
for spiral galaxies is based on their total ($i.e.$ spheroid and disk) luminosities.
Since we are interested in the relative number of globular clusters per unit 
spheroid lumniosity, the contribution of the disk to the overall 
luminosity of the spiral should be removed, as originally suggested 
by Harris (1981). From the catalog of Harris (1996a),
we have selected all giant elliptical and spiral galaxies having measured
specific frequencies. Two spirals studied recently
by Kissler-Patig et al. (1999) have also been included. Only spirals having
Hubble types between Sa and Sc have been considered,
since the difficulties involved in deriving
spheroid luminosities for later Hubble types become severe.

For each spiral, we calculate the specific frequency by combining the
total number of globular clusters taken directly from 
Harris (1996a) or Kissler-Patig et al. (1999). The spheroid luminosities
have been calculated using the ``bulge-to-disk" ratios given in the literature.
For those galaxies lacking published bulge-disk decompositions,
the mean ratio for the appropriate Hubble type given
in Simien \& de Vaucouleurs (1985) was used instead. For the 11 spirals,
which span the range Sa to Sc,
we find a mean specific frequency of ${\overline{S_n}} = 3.8\pm2.9$ measured 
with respect to their spheroids.

The mean specific frequency of giant elliptical galaxies in the Harris (1996a) catalog
is ${\overline{S_n}} = 5.2\pm3.2$. This value is indistinguishable
from that given above for spiral spheroids, although is probably an overestimate since it
includes a small number of ``high-$S_n$'' galaxies
located in the cores of the Virgo and Fornax clusters. The spirals, by contrast,
are almost invariably located in loose groups. In order to gauge the possible
importance of local environment,
we have calculated for each object the local galaxy density, $\rho_0$, 
using the number of galaxies brighter than $M_B = -16$ contained
within a shell of radius 0.5 Mpc using the Nearby
Galaxy Catalog of Tully (1988).
The 29 giant elliptical galaxies have 
$0.1 \lae \rho_0 \lae 9.2$ Mpc$^{-3}$, whereas the 11 spiral galaxies 
span the range $0.0 \lae \rho_0 \lae 3.7$ Mpc$^{-3}$. A total of 17 giant 
elliptical galaxies have $\rho_0 \lae 3.7$,
and are thus located in environments of comparable density. The
mean specific frequency of these galaxies is ${\overline{S_n}} = 3.9\pm2.5$,
which is indistinguishable from that found for the spiral spheroids.

\subsection{Metallicity Distributions}

The dramatic recent increase in the number of galaxies having accurately measured
globular cluster metallicity distributions is due primarily to the
use of metallicity-sensitive color indices ($e.g.$, Ostrov, 
Geisler \& Forte 1993) and the excellent imaging capabilities of the Hubble
Space Telescope ($e.g.$, Neilsen \& Tsvatanov 1999; Gebhardt \& Kissler-Patig 1999;
Kundu 1999). These studies 
have demonstrated that the majority of giant elliptical galaxies
show evidence for the presence of multiple, chemically-distinct 
globular cluster populations.

This multi-modality is not unique to giant elliptical galaxies: the Galactic
globular cluster system has long been known to
show a bimodal distribution in metallicity, showing
distinct peaks at [Fe/H] $\simeq -1.6$ and $-0.6$.
(Morgan 1959; Kinman 1959; Zinn 1985). Analysis of the globular cluster system 
associated with M31 suggests that it too appears significantly
multi-modal (Ashman \& Bird 1993; Barmby et al. 1999).
Unfortunately, the sample of spiral galaxies having accurate
globular cluster metallicity distributions is limited
to just these two objects, although recent work has demonstrated that 
bimodal globular cluster metallicity distributions are observed in S0 galaxies 
as well (Kissler-Patig et al. 1997b; Kundu \& Whitmore 1998). At present,
the only galaxies which show no evidence of multi-modal globular cluster 
metallicity distributions are dE/dSph galaxies of low and intermediate luminosity 
($i.e.$, $M_V \gae -17$). Such objects contain only metal-poor clusters 
(see Figure 1 of C\^ot\'e et al. 1998).

\subsection{Spatial Distributions}

All globular cluster systems studied to date have radial profiles
which are similar to, or shallower than, that of the underlying galaxy light 
(Harris 1986; Kissler-Patig 1997; Durrell et al. 1996). As Harris (1991) points out,
no case has been found in which the globular clusters system is more centrally
concentrated than the galaxy itself. 

The well-studied Virgo giant elliptical galaxies M49 and M87 are known to contain 
two chemically-distinct globular cluster populations; in both
cases, the metal-rich clusters appear to be more centrally concentrated than their 
metal-poor counterparts (Geisler et al. 1996; Neilsen, Tsvatanov \& Ford 1998; Lee, 
Kim \& Geisler 1998). In the case of M49, Lee et al. (1997) report a 6-$\sigma$ difference 
in the measured density profile slopes for the metal-rich and metal-poor
globular clusters (although see McLaughlin 1999 for a dissenting view).
In addition, Lee et al. (1998) note that the metal-rich clusters trace the
underlying galaxy light in both radial profile and ellipticity, whereas the
metal-poor clusters comprise a more extended and spherically distributed population.

At the present time, the only spiral galaxy which has been studied in sufficient detail 
that it is possible to investigate reliably the separate distributions of metal-rich 
and metal-poor clusters is the Milky Way.
The Galactic globular cluster system as a whole is known to obey a 
three-dimensional distribution of the form ${\rho}_h \sim R_G^{-m_h}$
where $m_h \simeq 3.5$ (Harris 1976). This is in close agreement
with the values of $m_h \simeq 3.0 - 3.5$ derived from halo 
RR Lyrae and blue horizontal branch stars (Saha 1985; Preston, Schectman \& Beers 1991).
Harris (1976) noted, however, that no metal-rich clusters
are found beyond $R_G \sim 7$ kpc, and suggested that most of these clusters 
are associated with the central bulge of the Galaxy, consistent with the findings
of several recent studies (Minniti 1995; Barbuy, Bica \& Ortolani; 
C\^ot\'e 1999; cf. Zinn 1985). For comparison, 
the density profile of the Galactic bulge is roughly 
${\rho}_b \sim R_G^{-m_b}$ where $3.65 \lae m \lae 4.2$ 
(Terndrup 1988, Blanco \& Terndrup 1989). Moreover, the mean Galactocentric radius
of the metal-rich cluster sample discussed by C\^ot\'e (1999) is 
${\overline{R_G}} = 3.2\pm2.0$ kpc, which is similar to the effective radius 
of the Galactic bulge ($R_{e} = 2.7$ kpc; Gilmore, King \& van der Kruit 1990).

In summary, the Galactic globular cluster system, like those of
all giant elliptical galaxies studied to date, has
a spatial distributions which is comparable to, or slightly more extended,
than that of the underlying spheroid light.
Moreover, the metal-rich globular clusters are more centrally 
concentrated than their metal-poor counterparts.

\section{A Hierarchical Model for the Formation of the Galactic Spheroid}

Given these similarities, it is natural to ask if a single model can explain
the observed properties of globular clusters in both giant elliptical and spiral
galaxies. For instance, models which seek to explain the formation of
giant elliptical galaxies and their associated systems of metal-rich globular 
clusters in spiral-spiral mergers ($e.g.$, Schwiezer 1986; Ashman \& Zepf 1992) suffer 
from the obvious difficulty that the spirals themselves appear to show 
multi-modal globular clusters metallicity distributions.

In the hierarchical picture of C\^ot\'e et al (1998), the metal-rich globular 
clusters in giant elliptical galaxies represent the clusters initially associated
with the most massive proto-galactic fragment. By contrast, the metal-poor globular 
clusters now associated with the giant elliptical galaxy are identified as those which 
have been acquired during
the accretion and disruption of numerous dwarf galaxies and proto-galactic fragments
($i.e.$, faint systems which are known to contain only metal-poor clusters).
The relative number of metal-rich and metal-poor globular clusters in the giant 
elliptical galaxy is then assumed to reflect the luminosities
(or, alternatively, masses) of the dominant proto-galactic building block and 
the accreted population of smaller proto-galactic fragments.
It is important to note that this procedure neglects possible differences in 
the destruction rates of metal-rich and metal-poor clusters (see \S 3.6).
Note that both the total number of clusters, and the globular cluster 
specific frequency, are conserved during the mergers ($i.e.$, we assume that 
no new clusters are formed in the merger process).

Thus, given three ingredients, it is possible to simulate the evolution of the
globular cluster metallicity distribution of a specific galaxy:
(1) the luminosity (or mass) function of galaxies and proto-galactic fragments; 
(2) the number of globular clusters per unit fragment luminosity (or mass) and; 
(3) the dependence of mean globular clusters metallicity on fragment luminosity (or mass).
In the present case, we have an additional constraint on the 
formation of the Galactic spheroid: the metallicity distribution of 
individual halo field stars. Specifically, any acceptable simulation of the 
formation of the Galactic spheroid must reproduce not only the observed metallicity
distribution of globular clusters (particularly, the two distinct peaks) 
but also that of the halo field 
stars. This latter distribution peaks at roughly the same metallicity as the metal-poor
globular clusters, and yet includes extended metal-poor and metal-rich tails which are not 
evident in the cluster distribution (Laird et al. 1988).

The various model inputs are discussed in detail below, where we concentrate on the
specific case of the hierarchical formation of the Galactic spheroid.

\subsection{Stellar Metallicity-Luminosity Relation for Proto-Galactic Fragments}

Following Larson (1988), Zinn (1993) and Mateo (1996), we begin by assuming that the dwarf
galaxy population of the Local Group can be thought of as the surviving
building blocks of their parent galaxies. Since it has been known for some time
that the mean stellar metallicity of galaxies depends rather sensitively on their
total luminosity ($e.g.$, Davies et al. 1987; Brodie \& Huchra 1991), we expect
that the stellar metallicity-luminosity relation defined by these galaxies is
a reasonable first approximation of that which would be expected for the proto-Galactic fragments
{\sl at the present time}. In other words, these fragments are assumed to have
faded passively with time in the same manner as the presently observed dwarfs. It is,
however, important to bear in mind that many Local Group dSph and dE galaxies have clearly
managed to form stars at intermediate epochs; this would not be the case for any 
proto-Galactic fragments which were disrupted and depleted of gas at very early times.

In the upper panel of Figure 1, the filled circles show the dependence of mean 
stellar metallicity on galaxy luminosity 
for 28 nearby dSph, dE and ``dSph/dIrr transition" galaxies
(Durrell et al. 1996ab; Mateo 1998; Caldwell et al. 1998; C\^ot\'e, Oke \& Cohen 1999).
The best-fit linear relation is given by
$${\overline{{\rm{[Fe/H]}}}_*} = -3.43(\pm0.14) - 0.157(\pm0.012)M_V \eqno{(2)}$$
which implies $L \propto Z^{2.54\pm0.19}$. For comparison, the dotted line 
indicates the scaling relation, $L \propto Z^{2.7}$, which is expected for dwarf 
galaxies which form in standard cold-dark-matter scenarios (Dekel \& Silk 1986).
Note the arbitrary metallicity zeropoint of the latter relation.
We conclude that equation (2) is a reasonable representation 
of the stellar luminosity-metallicity relation of surviving proto-Galactic fragments.
Although it has not been included in the fit,
the filled triangle in this figure indicates the location of the Galactic bulge
(McWilliam \& Rich 1994).
{\sl Note that its location is consistent with the extrapolation of the fitted 
relation for dwarf galaxies}.

Although the mean stellar metallicity in dSph and dE galaxies depends rather 
sensitively on total luminosity (and, presumably, total mass), there is now 
unmistakable evidence for sizeable abundance spreads within individual objects. 
The standard deviations in [Fe/H], assuming a Gaussian distribution of abundances, 
for these galaxies are shown in the lower panel of Figure 1. The data are 
taken mainly from the catalog of Mateo (1998), and have been supplemented with a few 
recent results on the M31 dSph system.
The mean value of ${\sigma}({\rm{[Fe/H]}_*}) = 0.36\pm0.11$ dex is indicated
by the upper dashed line. Unlike the mean metallicity, the dispersion in metallicity
depends weakly, or not at all, on total luminosity (C\^ot\'e, Oke \& Cohen 1999). 
The possible implications of this result are discussed in the following section.

\subsection{Globular Cluster Metallicity-Luminosity Relation for Proto-Galactic Fragments}

The open circles in the upper panel of Figure 1 show the mean metallicity of globular clusters
as a function of total galaxy luminosity for dSph and dE galaxies in the Local
Group, M81 and Virgo (see, $e.g.$, C\^ot\'e et al. 1998 and references therein). The 
best-fit linear relation is given by
$${\overline{{\rm{[Fe/H]}}}_{\rm GC}} = -3.79(\pm0.53) - 0.141(\pm0.033)M_V. \eqno{(3)}$$
The corresponding relation, $L \propto Z^{2.83\pm0.66}$, is, like the stellar
metallicity-luminosity relation, in excellent agreement with the predictions
of Dekel \& Silk (1986). It is also indistinguishable from that implied by equation (2)
with the notable exception of a ${\Delta}$[Fe/H] $\sim$ 0.6 dex offset between
the clusters and stars (in the sense that the clusters are a factor
of $\sim$ 4 more metal poor). 
For comparison, the open triangle indicates 
the mean metallicity of the metal-rich Galactic globular clusters (which has not
been included in the fit). The difference in metallicity between the bulge 
stars and metal-rich globular clusters is roughly ${\Delta}$[Fe/H] = 0.35$\pm$0.25 dex:
$i.e.$, smaller than, but consistent with, the metallicity offset seen in
the dwarf galaxies.

The lower panel of Figure 1 shows the intrinsic dispersions in metallicity 
for globular clusters in these galaxies, plotted against absolute visual magnitude
of the host galaxy. The mean value,
${\sigma}({\rm [Fe/H]}_{\rm GC}) = 0.30\pm0.11$ dex,
is indicated by the lower dashed line. To within the errors, this dispersion is 
the same as that of the stars and, similarly, shows no obvious trend with luminosity.

The theoretical metallicity-luminosity relation of Dekel \& Silk (1986) is based
on the key assumptions that these objects originated as gaseous protogalaxies embedded
in dominant dark-matter halos whose chemical enrichment was dictated by 
enrichment from massive stars and gas loss via supernovae-driven winds.
In such a scenario, the chemical evolution is approximated by the so-called ``Simple
Model" of chemical evolution (Searle \& Sargent 1972; Pagel \& Patchett 1975; 
Hartwick 1976). The success of this model in reproducing the observed metallicity-luminosity 
relations shown in Figure 1 suggests that it may also provide a convenient
representation of the metallicity distribution internal to each proto-Galactic 
fragment. This success is all the more remarkable in view of the
fact that many of the galaxies shown in Figure 1 exhibit incontrovertible
evidence for multi star-formation bursts, whereas the Dekel and Silk (1986)
model was framed within the context of a single star-fromation event.

For a homogeneous proto-Galactic gas cloud having zero initial metallicity and 
a yield, $y$, the metallicity distribution at the end of gas exhaustion takes the form:
$$df/dz \propto y^{-1}\exp{(-z/y)}. \eqno{(4)}$$
This distribution follows from the usual assumptions of the Simple Model: $i.e.$,
the initial mass function is constant in time, and the protogalactic
cloud experiences recycling of heavy elements from massive stars whose lifetimes
are short compared to the free-fall timescale of the gas cloud. In an attempt to
explain the lower metallicities of halo globular clusters
compared to the Galactic disk, Hartwick (1976)
defined an effective yield, $y_e$, given by the relation
$y_e = y/(1+c)$ where $c$ is a parameter related to the rate at which gas is
lost from the system (via, for example, supernovae-driven winds). Thus, in this
picture, the effective yield of each fragment or gas cloud is determined by
its overall mass (see, $e.g.$, Binney \& Merrifield 1998).

The upper panel of Figure 2 shows the stellar metallicity distribution predicted
by equation (4) for a galaxy having $M_V = -15$, where we have assumed that the 
effective yield is equal to the mean stellar metallicity predicted by 
equation (2).\altaffilmark{3}\altaffiltext{3}{Strictly speaking, $y_e$ refers to 
the {\sl mode} of the metallicity distribution given by equation (4).}
The lower panel shows the expected distribution for globular clusters assuming 
that the effective yield is given by equation (3). These distributions have, of 
course, identical shapes and full-widths at half-maxima ($\Gamma \simeq 1.0$ 
dex, corresponding to $\sigma \sim 0.44$ dex), but they are offset by
${\Delta}$[Fe/H] $\simeq$ 0.6 dex.
Based on the success of the Dekel \& Silk (1986) metallicity-luminosity 
relation, and the roughly constant spread in metallicity exhibited 
by both stars and clusters in these galaxies, we suggest that equation (4)
is a reasonable first approximation of the stellar and globular cluster metallicity 
distributions of dwarf galaxies and proto-Galactic fragments. Refinements to
the Simple Model, such as the inclusion of possible gas inflow 
and outflow, tend to produce narrower metallicity distributions ($e.g.$,
Gilmore, King \& van der Kruit 1990).

\subsection{The Luminosity and Mass Functions of Proto-Galactic Fragments}

As in C\^ot\'e et al (1998), the initial galaxian luminosity function ($i.e.$, the 
luminosity distribution of proto-galactic fragments) is approximated by a Schechter function
$$dN \propto (L/L^*)^{\alpha} \exp{(-L/L^*)}dL \eqno{(5)}$$
where $L^*$ is a characteristic luminosity and 
$\alpha$ is an exponent which governs the relative number of faint and bright systems 
(Schechter 1976). For early-type systems in low-density environments such as the 
Local Group, $L^*_B \simeq 8.2\times10^9L_{B,{\odot}}$ assuming 
$H_0$ = 75 km s$^{-1}$ Mpc$^{-1}$ (Marzke et al. 1998).
Since our goal is to model the growth of the Galactic spheroid (which has 
$L_B \sim 3.9\times10^9L_{B,{\odot}}$; see \S 3.4), the above representation is
effectively a power-law distribution in luminosity:
$$dN \propto L^{\alpha} dL. \eqno{(6)}$$
This distribution is similar to the mass spectrum expected in some
hierarchical cosmologies. 
For instance, in cold-dark-matter models,
the index of the initial power spectrum is $-3 < n < -2$ on the scales of
dwarf galaxies, which leads to mass function of the form $N(M) \propto M^{-2}$
($e.g.$, Blanchard et al. 1992; Ferguson \& Bingelli 1994). Unfortunately, the 
transformation to a luminosity spectrum remains highly uncertain since it involves
several poorly understood processes such as gas cooling, star formation
and feedback from massive stars. If the cosmological models are correct, then the fact
that the galaxy luminosity function today is shallower than the primordial mass
spectrum suggests that there is not a one-to-one correspondence between
galaxy mass and luminosity, and/or that the luminosity function has been
modified over time, perhaps by mergers (see \S 5).

The maximum fragment luminosity is dictated by 
the requirement that the total spheroid luminosity must not exceed the
observed value. The faint-end cutoff is taken to be $L_V = 2.7\times10^5L_{V,{\odot}}$,
the luminosity of the faintest galaxy ($i.e.$, Draco) used to define the metallicity-luminosity
relation of proto-Galactic fragments (\S 3.1). Fragments having luminosities below
that of the Fornax dSph galaxy ($i.e.,$ $L_V \le 1.6\times10^7L_{V,{\odot}}$) 
are assumed to contribute no globular clusters, since this is the faintest galaxy
known to contain its own globular cluster system.

\subsection{Number of Globular Clusters Per Proto-Galactic Fragment}

An alternative representation of equation (1) is
$$N_{gc} = {\eta}L_V \eqno{(7)}$$
where $\eta = (1.2\times10^{-8})S_n$ clusters $L_{V,{\odot}}^{-1}$.
There is some recent evidence that specific frequency
may not be a linear function of luminosity. For instance, as discussed
in \S 2.1, most early-type giant galaxies and spiral spheroids have 
globular cluster specific frequencies of $S_n \sim 4$, whereas the dSph and dE 
galaxies shown in Figure 1 have a marginally higher mean specific frequency 
of $S_n = 8\pm3$.  From {\sl HST} imaging of dE galaxies in the Virgo cluster, Miller et al.
(1999) find $S_n = 3.1\pm0.5$, with little or no luminosity
dependence. On the other hand, they find $S_n = 6.5\pm1.2$ for nucleated Virgo
dE galaxies, and see clear evidence of
a trend for $S_n$ to increase with decreasing luminosity ($i.e.$, rising
from $S_n \sim 3$ at $M_V \sim -17$ to $S_n \sim 20$ at $M_V \sim -13.5$)
Based on their results, and on the $N_{gc}$-$L_V$ relations for early-type dwarf 
and giant galaxies given in Kissler-Patig et al. (1997a) and 
McLaughlin (1999), we express the initial number of globular clusters, 
$N_{gc}$, associated with each proto-Galactic fragment as
$$N_{gc} = {\eta}L_V^{\beta} \eqno{(8)}$$
where $L_V$ is in solar units. For $L_V \le 2\times10^9L_{V,{\odot}}$, we
take $\beta = 0.8$ and $\eta \simeq 5\times10^{-6}$. Above this luminosity,
we assume $\beta = 1.1$ and $\eta \simeq 6\times10^{-9}$.

\subsection{Adopted Luminosities for the Galactic Bulge and Halo}

By definition, the luminosity of the Galactic spheroid, $L_V^s$, is given by the 
combined luminosities of the Galactic halo, $L_V^h$, and bulge, $L_V^b$. 
de Vaucouleurs \& Pence (1978) give $L_V^s = 4.7\times10^9L_{V,{\odot}}$ for the 
combined $R^{1/4}$
component of the Milky Way. This is considerably smaller than the value of
$L_V^s = 1.1\times10^{10}L_{V,{\odot}}$ found by Blanco \& Terndrup (1989).
In what follows, we shall adopt $L_V^s = 7.7\times10^9L_{V,{\odot}}$, 
which represents the mean of these two determinations.

Unfortunately, estimates of the {\sl separate} luminosities of the Galactic bulge and 
halo are uncertain due to the overlapping distributions of disk, halo and bulge 
stars in the inner
Galaxy (see, $e.g.$, Morrison 1996), the possible presence of a metallicity gradient in
the bulge (Minniti et al. 1995), and the unknown shape of the halo density profile in the inner
few kiloparsecs. In what follows, we adopt a bulge luminosity of 
$L_V^b = 5\times10^9L_{V,{\odot}}$ (Dwek et al. 1995; Holtzmann et al. 1998)
which, when combined with the above value of 
$L_V^s$, gives $L_V^h = 2.7\times10^9L_{V,{\odot}}$. This estimate is 
consistent with that of Suntzeff, Kraft \& Kinman (1991) who
used the relative space densities of globular clusters and field RR Lyrae stars to derive 
a total halo luminosity of $5.8\times10^8 L_{V,{\odot}}$ over
the range $4 \le R_G \le 25$ kpc: $i.e.$, our halo luminosity is equivalent to theirs for 
inner and outer limits on the halo population of $R_G \simeq$ 0.5 and 125 kpc,
respectively.

\subsection{Dynamical Evolution and Globular Cluster Destruction}

In their study of the globular cluster systems of giant elliptical galaxies, C\^ot\'e et al. (1998) 
made the first-order assumption that the metal-rich and metal-poor clusters have suffered
equal rates of destruction through dynamical processes. However, given the evidence for different
spatial distributions among the metal-rich and metal-poor sub-systems, this assumption
may be not valid since more rapid erosion is expected in the denser environments. 
As it seems inescapable that the Galactic globular clusters system 
has been depleted by dynamical processes (Ostriker, Spitzer \& Chevalier 1972; Tremaine 1974;
Fall \& Rees 1985; Aguilar, Hut \& Ostriker 1988; Gnedin \& Ostriker 1997; 
Murali \& Weinberg 1997) and that the likelihood of disruption for a given globular cluster 
depends sensitively on its orbit, erosive effects are expected to be more severe for the 
centrally-concentrated, metal-rich globular cluster system. 

We have attempted to incorporate the effects of dynamical evolution on the simulated globular
cluster metallicity distributions by adopting the results of the Fokker-Planck
calculations of Murali \& Weinberg (1997). These calculations include the combined effects
of relaxation, tidal heating and binary heating. The upper panel of Figure 3 shows
the initial and final cumulative radial distributions for Milky  Way spheroid 
globular clusters taken directly from Murali \& Weinberg (1997). In the lower panel, we have plotted
the ratio of the derivatives of these two curves, which we
take as a rough estimate of the ``survival probability", $P_S$, for a typical
globular clusters orbiting in the Galactic potential for a Hubble Time.
For $R_G >$ 30 kpc, the upper limit on Galactocentric radius considered by Murali \& 
Weinberg (1997), we take the $P_S(R_G) \equiv 1$. The minimum Galactocentric 
radius used in the simulations is $R_G = 0.8$ kpc. 

The presently observed density profiles of the (collisionless) bulge and halo field
star populations are then used to assign randomly an initial Galactocentric radius to each 
globular cluster. In other words, following Harris (1976),
Minniti (1995), Barbuy, Bica \& Ortolani (1998) and C\^ot\'e (1999), we associate the 
bulk of the metal-rich cluster population with the Galactic bulge, and not the thick-disk 
(Zinn 1985; Armandroff \& Zinn 1989). The Murali \& Weinberg (1997) survival probabilities
are then used to decide, on a cluster-by-cluster basis, which objects should be kept in the 
sample and which should discarded as likely candidates for disruption. The adopted 
density profile for the halo is $\rho_h (R_G) \propto R_G^{-3.5}$,
(Saha 1985; Preston, Schectman \& Beers 1991) while the bulge density profile 
is taken to be $\rho_b (R_G) \propto R_G^{-4.0}$ (Terndrup 1988; Blanco \& 
Terndrup 1990; Frogel et al. 1990).

Since the calculations of Murali \& Weinberg (1997) assume a fixed Galactic potential,
they may not be strictly appropriate for a model in which the Galactic spheroid is 
assembled from a collection of distinct proto-Galactic fragments. Nevertheless, they should 
at least provide a qualitative description of the dynamical evolution of the separate 
globular cluster systems since, in this scheme, dynamical erosion alters the overall number 
of metal-rich and metal-poor clusters but does not change the shape of their 
metallicity distributions.

\section{Comparison of the Observed and Simulated Metallicity Distributions}

\subsection{Methodology}

The algorithm used to generate the simulated metallicity distributions is based on that 
described in C\^ot\'e et al. (1998). The reader is referred to that paper for
a detailed disccusion of the model assumptions. Here, we give only a brief description 
of the model as it is applied to the specific case of the Milky Way.

The first step in the simulations is to generate a metal-rich system of bulge field
stars and metal-rich globular clusters by combining the adopted bulge luminosity 
with metallicity-luminosity relations given by equations (2) and (3), and by assuming
that the internal metallicity distribution of the bulge, like those all other 
proto-Galactic fragments, is accurately represented by equation (4). This procedure is
then repeated for additional proto-Galactic fragments, each drawn at random from the
luminosity distribution given by equation (5), until the combined luminosity 
is equal to observed luminosity of the Galactic spheroid. The
number of globular clusters belonging initially to each fragment is calculated using 
equation (8), while the relative number of stars contributed by the various fragments are given
simply by their luminosities. For globular cluster system, we approximate dynamical
evolution on a cluster-by-cluster basis using the Monte-Carlo approach discussed
in \S 3.6. The simulations are performed for a wide range in the adopted power-law 
exponent of the proto-Galactic luminosity function. A comparison of the observed and simulated
metallicity distributions for both the globular clusters and halo field stars 
is then used to decide which luminosity functions produce acceptable agreement. As in
C\^ot\'e et al. (1998), we assume that the proto-Galactic fragments have equal 
merger probabilities; the reader is referred to that paper for a discussion of the
possible effects of dynamical friction on the simulations.

\subsection{Globular Cluster Metallicity Distributions}

Perhaps the most noteworthy feature of the simulations presented here is the diversity of the 
end-products: the simulated globular cluster metallicity distributions 
show a wide range in appearance, ranging from unimodal distributions to more complex 
ones having multiple distinct peaks. This 
diversity is not unexpected in a stochastic process such as galaxy formation, and differs from the 
predominantly bimodal globular cluster metallicity distributions found previously for giant 
elliptical galaxies (C\^ot\'e et al. 1998) for two simple reasons. First, by virtue of the
globular cluster metallicity-luminosity relation and the modest luminosity
of the Galactic bulge, the mean metallicity of the metal-rich
clusters is not as widely separated from that of the metal-poor component.
Second, the high luminosities of giant elliptical galaxies permit the accretion of 
correspondingly more luminous proto-galactic fragments or galaxies, meaning
that the exponential cutoff in the luminosity distribution given by equation (5)
imposes a sharp cutoff on the metal-rich side of the distribution of 
globular clusters arising in proto-galactic fragments. Such a cutoff, 
which serves to dilineate the globular clusters of the dominant 
proto-galactic fragment from those of the other fragments, does not apply
in the case of the Galactic spheroid since $L_V^s \lae L_V^*$, as
discussed in \S 3.3 and 3.5.

The principal conclusions drawn from these simulations can be
summarized as follows: (1) the bulge, as the dominant
proto-Galactic building block, is observed to have the most metal-rich globular
clusters system by virtue of the globular cluster metallicity-luminosity relation;
(2) the bulge contributes roughly {\sl twice} the number of globular clusters initially as do the
combined halo progenitors; (3) the more centrally concentrated metal-rich
globular cluster system has been preferentially eroded by dynamical effects;
(4) the metal-poor cluster system exhibits a much wider range in its observed properties
although the simulations reveal that $\alpha \sim -2$ produces the closest match
to the observed distribution, particularly the peak at [Fe/H] $\sim -1.6$; and
(5) the metal-poor clusters, being more spatially extended than their metal-rich
counterparts, have undergone less severe dynamical erosion.

The upper panel of Figure 4 shows a single simulation of globular cluster metallicity distribution,
{\sl specifically chosen to match roughly the basic properties of the Galactic
globular cluster system.} The actual distribution, based on a sample of
133 clusters having measured metallicities (Harris 1996b), is
indicated by the open circles. The two dotted curves show the initial 
distributions of bulge and halo globular clusters, while the dashed curves
indicate the same distributions after including the effects dynamical erosion.
The spheroid in this simulation has $L_V^s = 7.4\times10^9L_{V,{\odot}}$
and contains 149 surviving globular clusters, 
descended from an original population of 429. In this particular simulation, the halo 
was assembled from a total of 1309 proto-Galactic fragments having a combined luminosity 
of $L_V^h = 2.7\times10^9L_{V,{\odot}}$. 
The final specific frequency of the Galactic spheroid is $S_n = 1.6$,
which is identical to the observed value.

\subsection{Halo Star Metallicity Distributions}

The solid curve in the lower panel of Figure 4 shows the halo star metallicity distribution 
for the simulation described above. Open circles indicate the actual metallicity distribution of 
372 kinematically selected halo field stars according to Ryan \& Norris (1991). 
Both the simulated and observed distributions show maxima in the range $-1.7 \lae$ [Fe/H] $\lae -1.5$ 
and broad wings extending to lower and higher metallicities.
These extended tails are more pronouced in the field star distribution than in 
the cluster metallicity distribution. The significance of these 
differences has always remained somewhat questionable due to the
finite size of the Galactic cluster system ($i.e.$, the discrepancy at the metal-poor 
and metal-rich ends can removed by adding only four and six clusters, respectively;
Laird et al. 1988).

Nevertheless, such differences are often seen in the simulations
described here and their origin can be understood as follows.
The extended metal-poor tail in the field star distribution is populated exclusively by 
stars formed in the faintest and most metal-deficient proto-Galactic fragments:
$i.e.$, having mean metallicities of [Fe/H] $\sim -2.1$ (see Figure 5). 
Such a tail is slightly less evident in the globular cluster distribution 
since only proto-Galactic
fragments having $L_V \gae 1.6\times10^7L_{V,{\odot}}$ contribute clusters to the
spheroid, limiting the mean metallicity of these clusters to [Fe/H] $\sim -1.9$.
More significantly, the comparatively large number of faint proto-Galactic
fragments incorporated into the halo ensures that the metal-poor tail of simulated halo 
metallicity distribution is well populated, in constrast to that of the globular
cluster distribution (which, unlike the field star distribution, is further 
depleted by dynamical effects).
At the metal-rich end, the small number of luminous proto-Galactic fragments
incorporated into the spheroid contribute significant numbers of both globular clusters
and field stars; as discussed in \S 3.2, these stars will be systematically 
$\sim$ 0.6 dex more metal-rich than the associated clusters 
and, consequently, will produce a metal-rich tail which will not be seen
in the globular cluster metallicity distribution.

For comparison, the dotted curve in the lower panel of Figure 4 shows the prediction 
of the Simple Model for an effective yield of
${\log_{10}{y_e}} = -1.6$, an {\sl arbitrary} value chosen by Ryan \& Norris (1991) to 
give the closest match to the observed distribution.
As noted by both Laird et al. (1988) and Ryan \& Norris (1991), at high metallicities 
the Simple Model shows poor agreement with the actual distrubution. While the significance
of this discrepancy is unclear due to possible contamination by metal-rich
disk stars, we note that the simulations
show significant numbers of stars having [Fe/H] $> -1$ whose origin
can be traced to the largest proto-Galactic fragments incorporated 
in the spheroid. The fraction of such stars in the simulations, however,
is somewhat larger than that seen in the distribution of
Ryan \& Norris (1991).

A consistency check on the radial distribution of the simulated globular cluster systems is 
shown in Figure 6. The upper and lower panels indicate histograms of Galactocentric radii for 
the actual and simulated globular clusters systems. In the latter case, we plot the
radial profiles before and after dynamical effects. Initially, the simulated
globular clusters follow a profile given by the adopted halo and bulge density laws, shown as the
solid and dashed lines in the upper panel. Afterwards, the surviving globular clusters have 
a radial distribution similar to that of the observed globular clusters, including the same 
flattening of the profile in the central $R_G \sim$ 3-5 kpc. For the distant metal-poor
globular clusters, the initial distribution is relatively unaltered.

\subsection{How Fragmented was the Proto-Galactic Spheroid?}

Figure 7 illustrates some properties of the proto-Galactic fragments 
from which the Galactic spheroid was assembled. In the upper panel, we show 
the luminosity distribution of proto-Galactic fragments found in the 
above simulation. The location of the proto-bulge is 
indicated by the vertical arrow, while the dashed line indicates a power-law 
luminosity function having slope $\alpha = -2$. The dotted line in the lower 
panel shows the cumulative luminosity distribution of these
same proto-Galactic fragments. Given the steep luminosity function, 
the vast majority of the proto-Galactic fragments are, as expected, low-luminosity systems.
Unfortunately, the conversion from luminosity to mass for these fragments is 
highly uncertain due their unknown mass-to-light ratios.
For illustrative purposes, two different mass 
distributions are shown in the lower panel of Figure 7. In the first case, we have assumed
$M/L_V = 2$ in solar units. This would be expected for an old stellar population
whose dark matter content can be understood purely in terms of normal stellar remnants,
as is the case for globular clusters
(Gunn \& Griffin 1979; Pryor \& Meylan 1993). In the second case, we have assumed $M/L_V = 2$
and a universal dark halo mass of $M \simeq 2.0\times10^7M_{\odot}$, as suggested
by studies of the internal kinematics Local Group dwarf galaxies (Mateo 1993).

These simulations suggest that proto-Galactic spheroid was highly fragmented into 
numerous distinct, chemically-isolated fragments. For $\alpha = -2$, the total number of fragments
is $N_{PGF} \sim$ 1-2$\times10^3$,
which follows directly from the assumed power-law index for the luminosity function, the
total luminosity of the Galactic spheroid and the adopted faint-end cutoff of the proto-Galactic 
luminosity function. The majority of these proto-Galactic fragments are low-luminosity
systems, with roughly 95\% of the fragments having $L_V \lae 1.6\times10^7L_{V,{\odot}}$
($i.e.$, the present day luminosity of the Fornax dSph galaxy).
As a whole, these faint fragments contribute nearly half of the total halo luminosity but,
by virtue of their low luminosity, none of its globular clusters. This 
is evident in the upper panel of Figure 5 which shows, 
for one simulation, the metallicities of halo
stars and globular clusters plotted against the luminosity of the proto-Galactic
fragment in which they originated. Although the globular clusters and field 
stars have similar mean metallicities, the field star distribution extends
to both higher and lower metallicities, as evident in the lower panel of Figure 5.

Given the large number of proto-Galactic fragments required by hierarchical formation 
models to match the observed metallicity distributions, it is interesting to 
consider the possible implications for the 
mass budget of the Galaxy. If it is assumed that each fragment consists of
a luminous component having $M/L_V = 2$ which is embedded in a constant mass dark matter
halo as described above, then for the adopted cutoff of $L_V = 2.7\times10^5L_{V,{\odot}}$, 
the total mass is $M \simeq 3\times10^{10}M_{\odot}$. This is much lower than the total
Galactic mass of $M \sim$ 3-9$\times10^{11}M_{\odot}$ (Zaritsky 1989; Kochanek 1996), 
suggesting that the proto-Galactic fragments alone cannot account for the dark matter 
content of the Milky Way.

\subsection{A Second Example: The M31 Spheroid}

M31 has traditionally presented a challenge to models of halo formation since it
is difficult to understand why its halo stars are, on average, four
times more metal-rich than its globular clusters (Mould \& Kristian 1986; Brodie \& 
Huchra 1991). This difference is all the more puzzling in light of the fact
that these components in the Milky Way --- the other large Local Group 
spiral --- have nearly identical mean metallicities.

A possible explanation of this difference is shown in Figure 8.
In the upper panel, we compare the observed and simulated globular cluster
metallicity distribution for M31; the botton panel shows the observed and
simulated metallicity distributions for M31 halo stars. The data are taken
from Barmby et al. (1999) and Holland et al. (1996), respectively.
No attempt has been made to include dynamical effects for the globular clusters
since there are no published calculations of the dynamical evolution of the M31
globular cluster system. While dynamical effects will influence the relative
numbers of globular clusters in the metal-rich and metal-poor populations,
the mean metallicities of the two components will be not be affected.
Following Walterbos \& Kenicutt (1988), we adopt 
$L_V^s = 9.8\times10^9L_{V,{\odot}}$ for M31 and assume 
an identical bulge-to-halo ratio as for the Milky Way.
Experiments indicate that $\alpha \sim -1.8$ most 
frequently produces the best match to the observed halo star distribution, 
particularly the peak at [Fe/H] $\simeq -0.7$.
The observed and simulated cluster distributions, meanwhile, 
have their maxima at [Fe/H] $\simeq -1.3$ due
to the fact that the assumed luminosity function of the proto-galactic fragments
is slightly more skewed toward higher luminosity fragments than was the case for 
the Milky Way.  On the other hand, the prominent metal-poor tail seen in the halo 
star distribution is less pronounced in the simulations.
As Holland et al. (1996) point out, the significance of this tail is unclear
since confusion between metal-poor red giant branch stars and metal-rich
asympototic giant branch stars becomes important at this level.

Figure 9 shows the dependence of halo star and globular cluster metallicity on
proto-galactic fragment luminosity. A comparison with Figure 7 reveals the clear 
differences between the simulated metallicity distribution for the Galaxy and M31.
In the case of M31, the proto-galactic fragments have a slightly flatter luminosity
function which results in a field star distribution whose mean metallicity
is several times higher than that of the associated globular clusters.

\section{Discussion and Implications}

The simulations presented here demonstrate that hierarchical models are able
to provide an excellent match
to the metallicity distributions of Galactic globular clusters and halo
field stars, {\sl provided the luminosity function of proto-Galactic fragments 
had the form $dN \propto L^{\alpha}dL$ with $\alpha \sim -2$}. 
Such a steep slope is in agreement with the predictions of 
semi-analytic/numerical models of hierarchical galaxy formation and the standard 
assumptions regarding
gas cooling in dark halos ($e.g.$, White \& Rees 1978; Blumenthal et al. 1984;
Kauffmann, White \& Guiderdoni 1993; Klypin et al. 1999; Moore et al. 1999). 
It is, however, strongly
inconsistent the presently observed luminosity function of the Local
Group and of the field galaxy population in general. While it is undoubtedly
true that the current census of Local Groups galaxies is incomplete at
the faint end, it is highly unlikely that the number of Local Group
dwarf galaxies has been underestimated by more than an order of magnitude. For 
instance, Pritchet
\& van den Bergh (1999) have shown that the Local Group luminosity function
closely resembles a Schechter function having $\alpha \simeq -1.1$ and that,
based on the current census of Local Group galaxies,
the probability of $\alpha < -1.3$ for a single Schechter function is 
less than one percent. 

Is it possible to reconcile these results? It is worth pointing out that this 
discrepancy is not unique to the simulations presented here, but is rather
a longstanding problem for semi-analytic and numerical hierarchical models. For instance, 
Klypin et al. (1999) noted, on the basis of high-resolution cosmological simulations
of the Local Group, that the number of low-mass, dark halos predicted by
the models exceeds the observed number of faint galaxies by nearly an order
of magnitude. The magnitude of the discrepancy 
is slightly less than the one found here, since
Klypin et al. (1999) predicted $\sim$ 300 halos within 1.5 Mpc of the Local Group
(compared to the observed number of $\sim$ 40 satellites), although the larger number of 
proto-Galactic fragments found in the present case may be a consequence of the different
mass/luminosity cutoffs. That is to say, the simulations of Klypin et al. (1999)
become incomplete below $v_c \sim 20$ km s$^{-1}$ whereas such low-mass systems
are explicitly included in our Monte-Carlo approach: $i.e.$, $v_c \sim 10$ km s$^{-1}$ 
at the faint end of the proto-Galactic luminosity function (Mateo 1998).

We suggest that the proto-galactic fragments discussed here are plausible
candidates for the low-mass, dark halos seen in the semi-analytic and
numerical models. The issue, however, is complicated by the fact that, in the
present case, the spheroid is identified with the
disrupted stellar components of these fragments, whereas $N$-body simulations
suggest that low-mass halos are relatively immune to the destructive effects of
the Galactic tidal field. If this association is correct, then some physical
mechanism is required to erase the dark matter substructure observed in the 
numerical models, such as tidal heating of halos on predominately radial
orbits (Moore et al. 1996; van den Bosch 1999) or impulsive heating during
rapid, halo-halo encounters (Moore et al. 1996).

Two possible explanations for the discrepancy between the predicted and observed
luminosity functions were discussed by Klypin et al. (1999):
high-velocity clouds (HVCs) and dark satellites. In the first case, the numerous 
HVCs which populate the Local Group are assumed to 
represent the observable counterparts of the lowest-mass dark halos (Blitz et al.
1999). Several properties of the HVCs, such as their large 
numbers (about 2500 in the Local Group; Stark et al. 1992), their low masses 
(typically $3\times10^7M_{\odot}$ of
neutral gas and roughly ten times this amount of dark matter; Blitz et al 1999) 
their presumed extragalactic nature, and their inferred high rate of accretion onto 
the Galaxy at early times (Blitz et al. 1999), make them attractive candidates for the
proto-Galactic fragments described here. Evidence for spatial and kinematic
connections between HVCs and at least some Local Group dwarf galaxies has recently 
been presented by Blitz \& Robishaw (1999) and C\^ot\'e et al. (2000).
However, the dSph and dE galaxies
which we have identified as the surviving proto-Galactic fragments have clearly
managed to convert much of their initial gas reservoir into stars, something 
which is not true for the majority of the HVCs. 

Klypin et al. (1999) also examined the possibilty that many low-mass dark halos
rapidly lost their gas due to supernovae-driven winds or an intergalactic 
photo-ionizing background.
Although these processes may be important for 
explaining the excess in the predicted number of dark halos over the number
observed, they cannot resolve the discrepancy found here.
If the hierarchical models are correct, then the very 
existence of the Galactic spheroid demonstrates that its constituent proto-Galactic 
fragments managed to form significant numbers of stars.

A related possibility is that the luminosity function of
proto-Galactic fragments depends sensitively on local environment. The 
steep slopes required by hierarchical formation models in the immediate 
vicinity of the proto-Galaxy might then be a consequence of pressure
confinement in high-density regions (Babul \& Rees 1992), biased dwarf
formation (West 1993; Ferguson \& Binggeli 1994) or some other mechanism
which enhances the efficiency of gas cooling in low-mass halos. 
Indeed, the remarkable diversity in the star formation histories of 
Local Group dwarfs ($e.g.$, Grebel 1999) provides {\it prima facie} evidence for 
the complexity of gas accretion, cooling and ejection in these objects.

Finally, the simulations presented here provide no direct constraints on the timescale of
spheroid assembly, but it is nevertheless possible to draw some general conclusions on the 
duration of spheroid assembly. While the existence of the disrupting Sagittarius dwarf 
galaxy (Ibata, Gilmore \& Irwin 1995) suggests
that the accretion process has continued up to the present day,
other arguments indicate that the majority of proto-Galactic fragments must 
have been incorporated into the spheroid at very early times.
First, the thinness of the Galactic disk, whose oldest stars are believed 
to be 10$^{+3}_{-1}$ Gyr old (Wood \& Oswalt 1998; Knox, Hawkins \& Hambly 1999), 
may indicate that the number of massive satellites accreted over its lifetime 
has been small (Toth \& Ostriker 1992; 
Moore et al. 1999).\altaffilmark{4}\altaffiltext{4}{Estimates of disk heating by 
infalling satellites are reduced if the disks are allowed to warp ($e.g.$,
Huang \& Carlberg 1997; Sellwood, Nelson \& Tremaine 1998) but, as Moore et al. (1999) have 
pointed out, such a mechanism is unlikely to be effective in the present case due to 
the large number of proto-Galactic fragments.}
Second, a majority of the Local Group dSph/dE galaxies contain young- and intermediate-age
stellar populations, whereas the fraction of such stars in the halo is
known to be small ($i.e., \lae$ 10\%; Unavane et al. 1998). 
However, it is important to bear in mind that the
proto-Galactic fragments, if accreted and disrupted at early times,
would not have had the opportunity to form stars over periods of time,
as did the Local Group dwarfs. In summary, the available evidence seems to favor 
an early, and relatively rapid, timescale for the assembly of the Galactic spheroid.

\section{Summary}

We have described a semi-empirical technique for simulating the chemical evolution
of the Galactic spheroid in hierarchical formation scenarios. The simulations 
include no gas dissipation, but assume instead that the bulk of star and cluster
formation occured within distinct, chemically-isolated proto-Galactic fragments
which were subsequently assembled into the Galactic spheroid. The chemical
enrichment of each proto-Galactic fragment is assumed to proceed in the manner
predicted by the Simple Model (Searle \& Sargent 1972; Pagel \& Patchett 1975;
Hartwick 1976). The effective yield of each fragment is 
determined empirically using the presently observed metallicity-luminosity
relations for stars and globular clusters belonging to nearby dSph and dE
galaxies.

In this picture, the bulge is identified as the dominant proto-Galactic building
block, and the metal-rich Galactic globular clusters as its associated cluster
system. This identification is supported by the observation that the metallicities 
of bulge field stars and the metal-rich Galactic globular clusters are 
consistent with the extrapolated metallicity-luminosity relations of dwarf
galaxies ($e.g.$ the smaller proto-Galactic fragments). By contrast, the
Galactic halo is identified as the disrupted remains of numerous, much smaller,
proto-Galactic fragments. 
A comparison between the observed and simulated metallicity distributions of Galactic
globular clusters and halo field stars shows good agreement, {\sl provided the 
luminosity function of proto-Galactic fragments has the form $dN \propto L^{{\alpha}}dL$ 
where $\alpha \sim -2$.} When combined with the observed 
luminosity of the Galactic halo, this steep slope implies that the proto-Galactic
spheroid was fragmented into $N_{PGF} \sim 10^3$ distinct star-forming
regions; the metal-poor Galactic globular clusters formed in the $\sim$ one dozen
most massive fragments, whereas the bulk of the halo field star population and, 
in particular, the most metal-deficient objects, originated in numerous
smaller fragments.

While these simulations provide independent support for semi-analytic and numerical 
models of hierarchical galaxy formation, they exacerbate the longstanding discrepancy 
between the observed and predicted number of nearby faint galaxies.

\acknowledgments
 
We thank Pauline Barmby, Sean Ryan and Steve Holland for providing the cluster and 
stellar metallicities shown in Figures 4 and 9. Thank also to the referee, Mario Mateo, 
for his many helpful suggestions.
P.C. acknowledges support provided by the Sherman M. Fairchild Foundation.
Additional support for this work was provided to R.O.M. by NASA through grant No. 
HF-0.096.01-97A 
from the Space Telescope Science Institute, which is operated by the 
Association of Universities for Research in Astronomy, Inc., under NASA contract NAS5-26555. 
M.J.W. acknowledges financial support from the Natural Sciences and Engineering
Research Council of Canada. 
DM is supported by the Chilean Fondecyt Project No.\ 01990440, and by 
the U. S. Department of Energy by Lawrence Livermore National Laboratory 
under Contract W-7405-Eng-48. This research has made use of the NASA/IPAC Extragalactic 
Database (NED) which is operated by the Jet Propulsion Laboratory, California Institute of 
Technology, under contract with the National Aeronautics and Space Administration.

\clearpage
 
\plotone{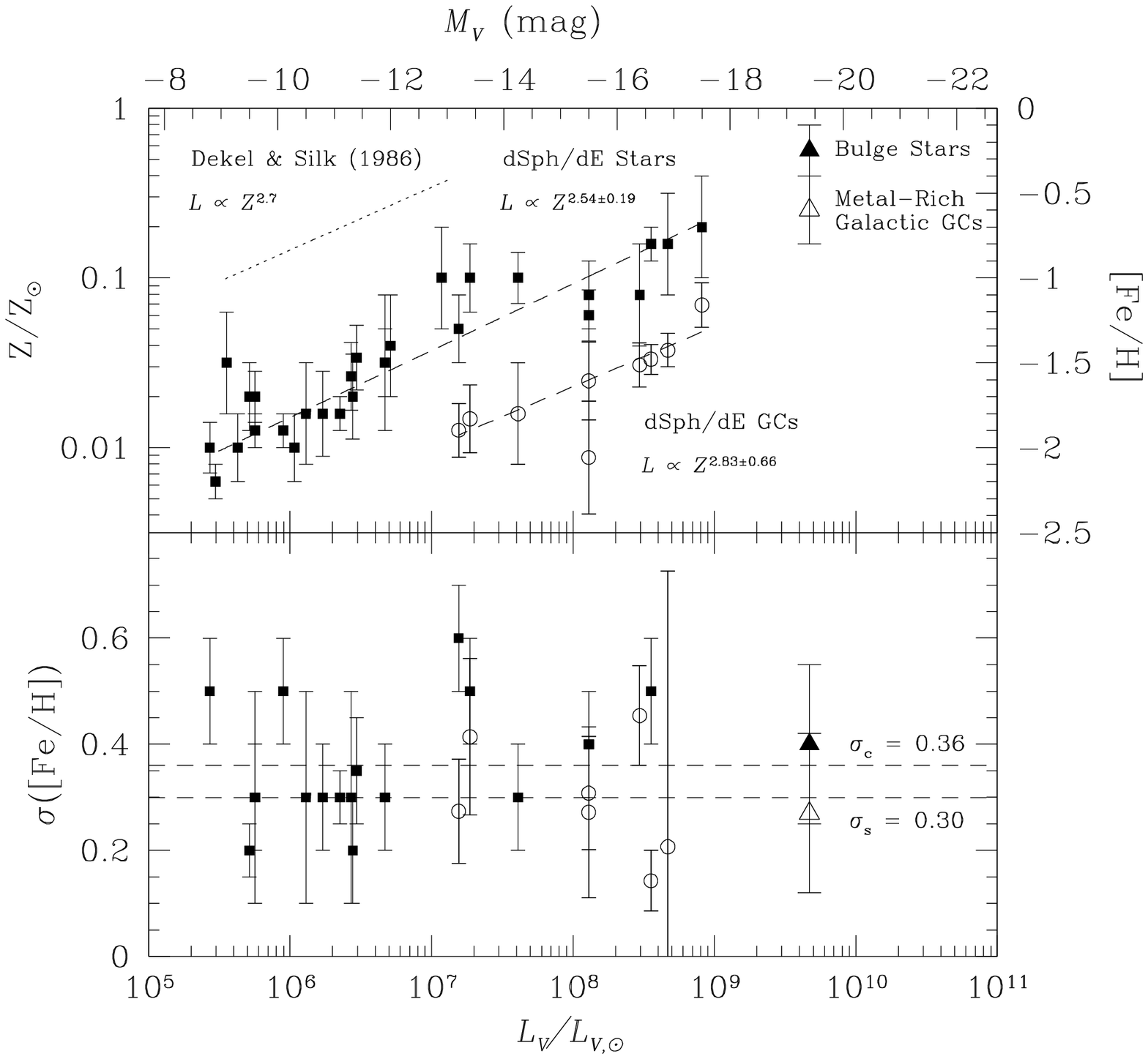}

\figcaption[mws.01.ps]{
(Upper Panel) Relation between total luminosity and the mean metallicity of stars 
(filled squares) and globular clusters (open circles) belonging to early-type dwarf
galaxies. The filled triangle indicates the mean metallicity of
Galactic bulge stars; the open triangle shows the mean metallicity of metal-rich 
globular clusters in the Milky Way.
The best-fit linear relations for the stars and clusters 
are indicated by the dashed lines while the solid line
indicates the predicted relation of Dekel \& Silk (1986), shifted
vertically by an arbitrary amount. 
This theoretical relation is indistinguishable from the empirical relations 
observed for both the stars and globular clusters.
(Lower Panel) Observed dispersion in metallicity for stars (squares) and 
globular clusters (circles) for the same sample. The filled
and open triangles indicate the observed dispersions for Galactic bulge stars
and metal-rich Galactic globular clusters, respectively. 
\label{fig1}}
\clearpage

\plotone{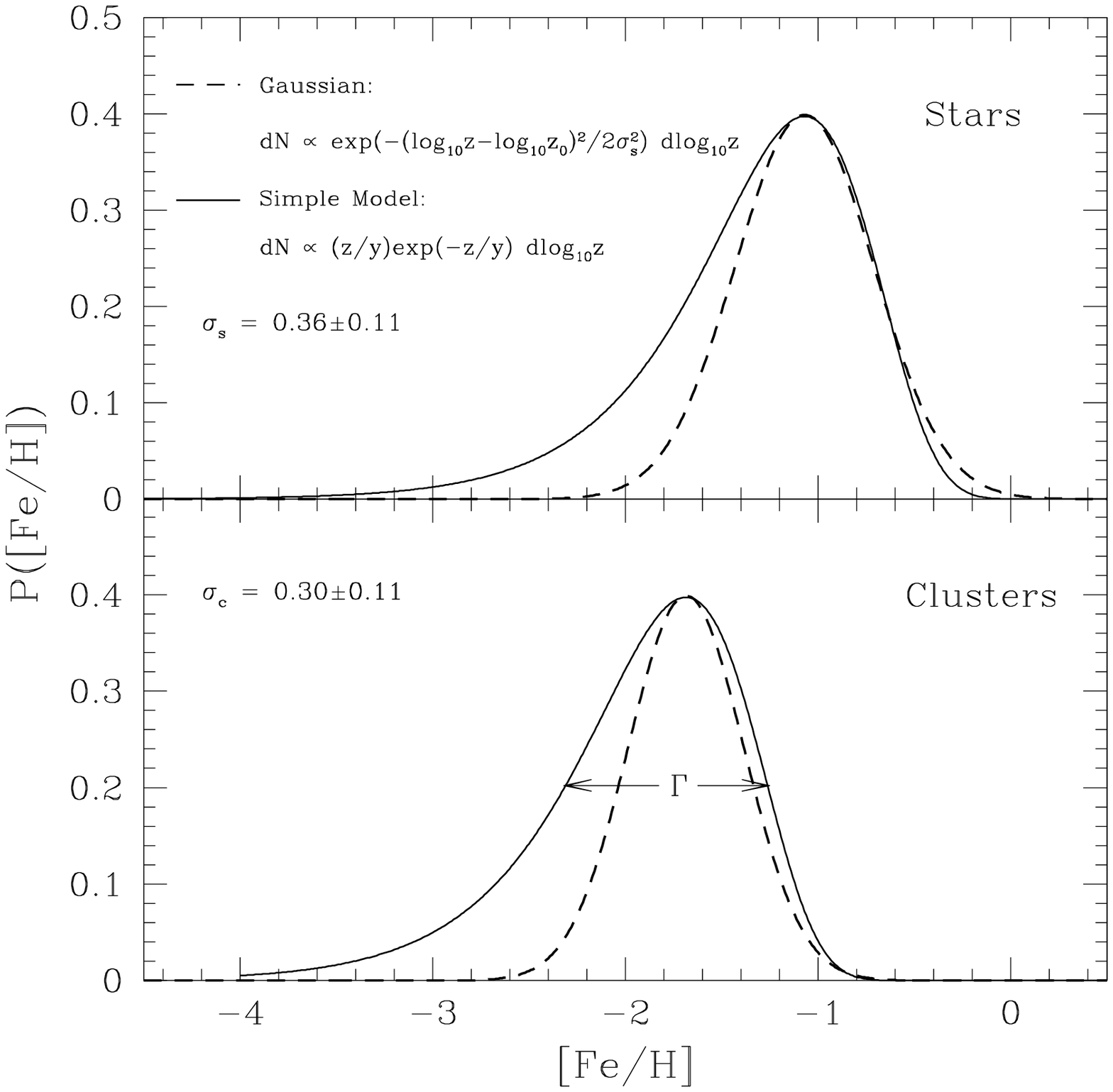}

\figcaption[mws.02.ps]{
(Upper Panel) Predicted internal stellar metallicity distribution according to the 
Simple Model for a proto-Galactic fragment having a present-day luminosity of 
$L_V = 8.2\times10^7L_{V,{\odot}}$ ($e.g.$, $M_V = -15$). The effective yield has 
been set equal to the mean
metallicity given by equation 2. For comparison, the dashed curve shows a 
Gaussian with dispersion $\sigma_s$([Fe/H]) = 0.36 and the same modal metallicity.
(Lower Panel) Predicted distribution of globular cluster metallicities based
on the Simple Model for this same proto-Galactic fragment. In this case, the effective 
yield is equal to the mean metallicity given by equation 3. As before, the dashed
curve shows a Gaussian with dispersion $\sigma_s$([Fe/H]) = 0.30 and the same 
modal metallicity.
\label{fig2}}

\clearpage

\plotone{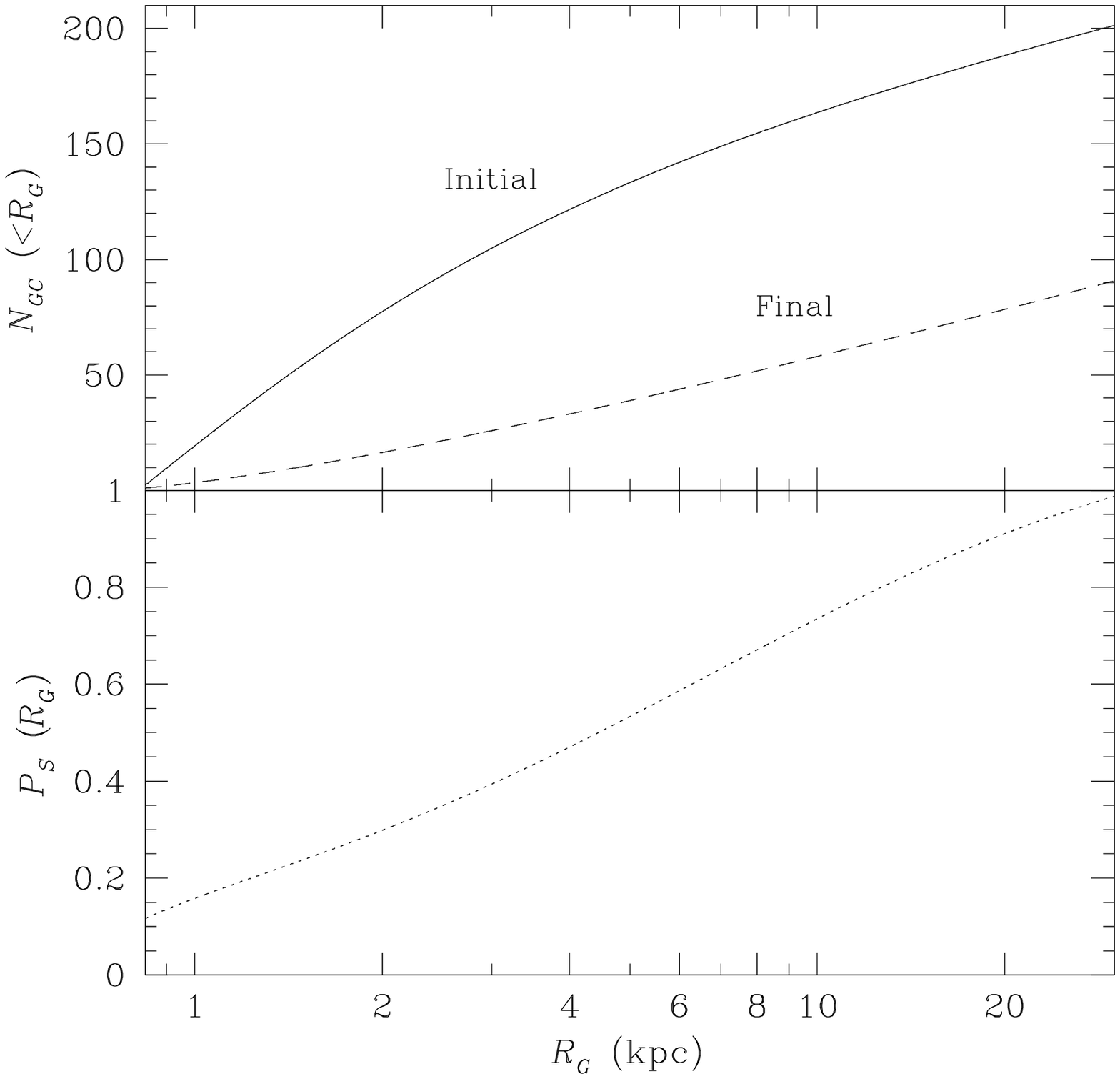}

\figcaption[mws.03.ps]{
(Upper Panel) Initial and final distributions for globular clusters belonging to the
Galactic spheroid according to the Fokker-Planck calculations of Murali
\& Weinberg (1997). (Lower Panel) Survival probability as a function of
galactocentric radius for globular clusters calculated from the cumulative
distributions shown above (see text for details). Clusters beyond $R_G = 30$ 
kpc are assumed to have survival probabilities of unity.
\label{fig3}}
\clearpage

\plotone{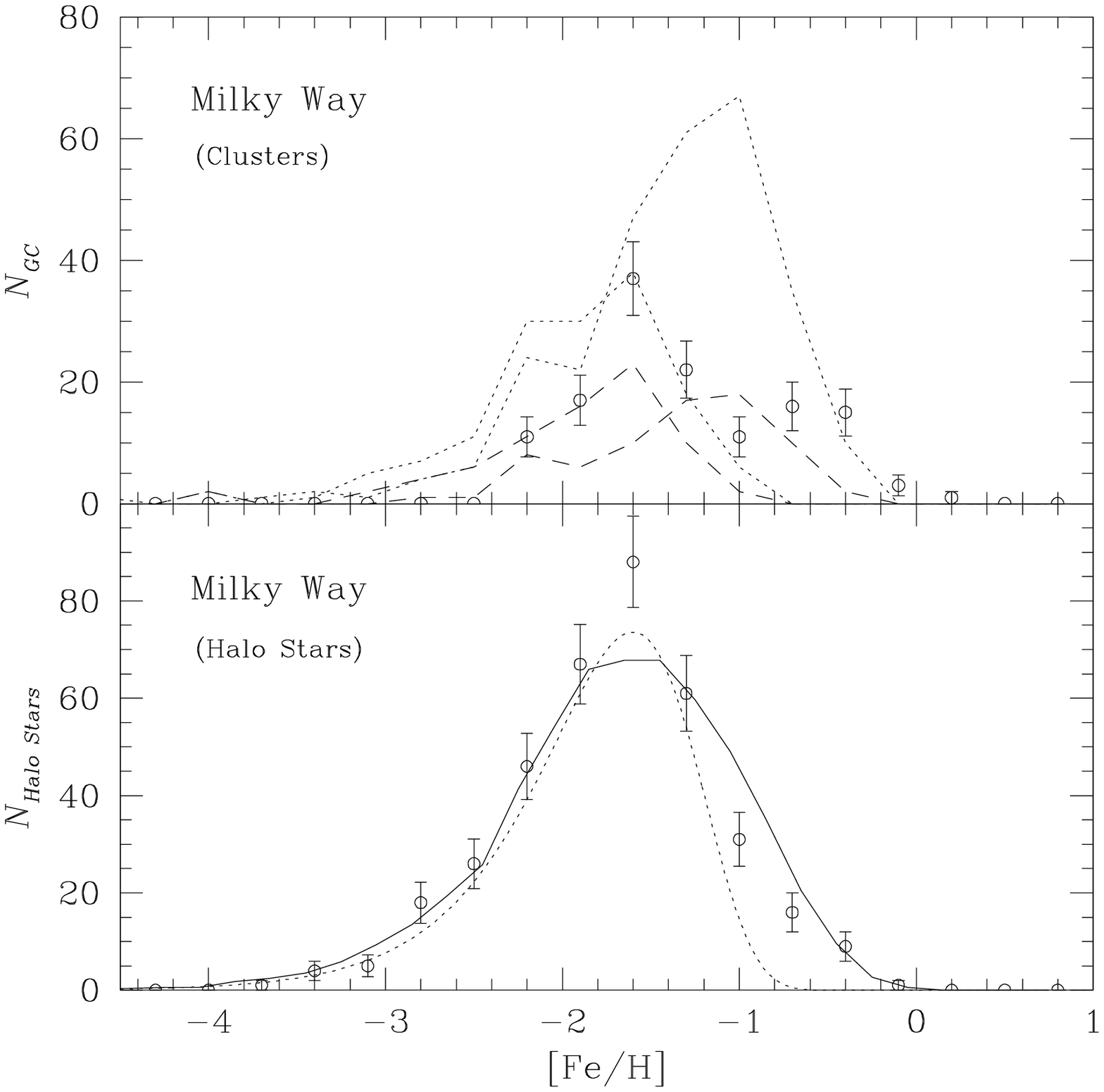}
\figcaption[mws.04.ps]{
(Upper Panel) One simulation of the Galactic globular cluster metallicity distribution, 
before (dotted curves) and after (dashed curves) the effects of dynamical 
evolution are included. The metal-rich and metal-poor components indicate the
respective globular cluster systems of the proto-bulge and proto-halo.
The actual distribution, based on data from Harris (1996b), is indicated 
by the open circles. 
(Lower Panel) Corresponding distribution of halo field star metallicities
based on this simulation (solid curve). The open circles show the halo metallicity 
distribution based on the data of Ryan \& Norris (1991).
The dotted line shows the prediction of the Simple Model for an effective yield
of ${\log{y_e}} = -1.6$.
\label{fig4}}

\clearpage
\plotone{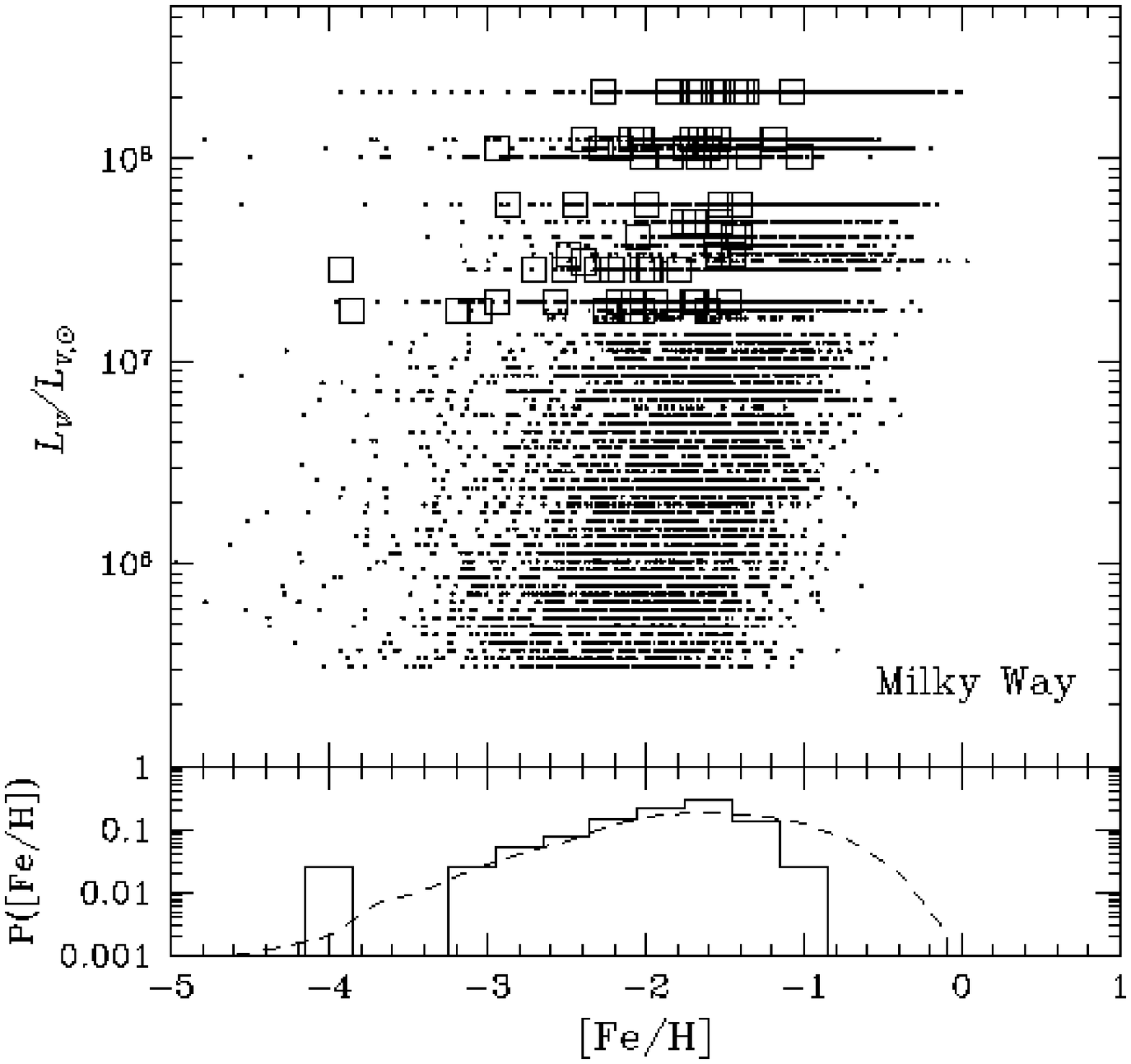}
\figcaption[mws.05.ps]{
(Upper Panel) Metallicities of globular clusters (squares) and halo field stars (dots)
plotted as a function of the luminosity of the proto-Galactic fragment in which they
originated.
(Lower Panel) Comparison of the metallicity distributions of globular clusters
(solid curve) and halo field stars (dashed curve). Although the two
samples have similar mean metallicities, the halo star population shows more
extended metal-poor and metal-rich tails.
\label{fig5}}

\clearpage
\plotone{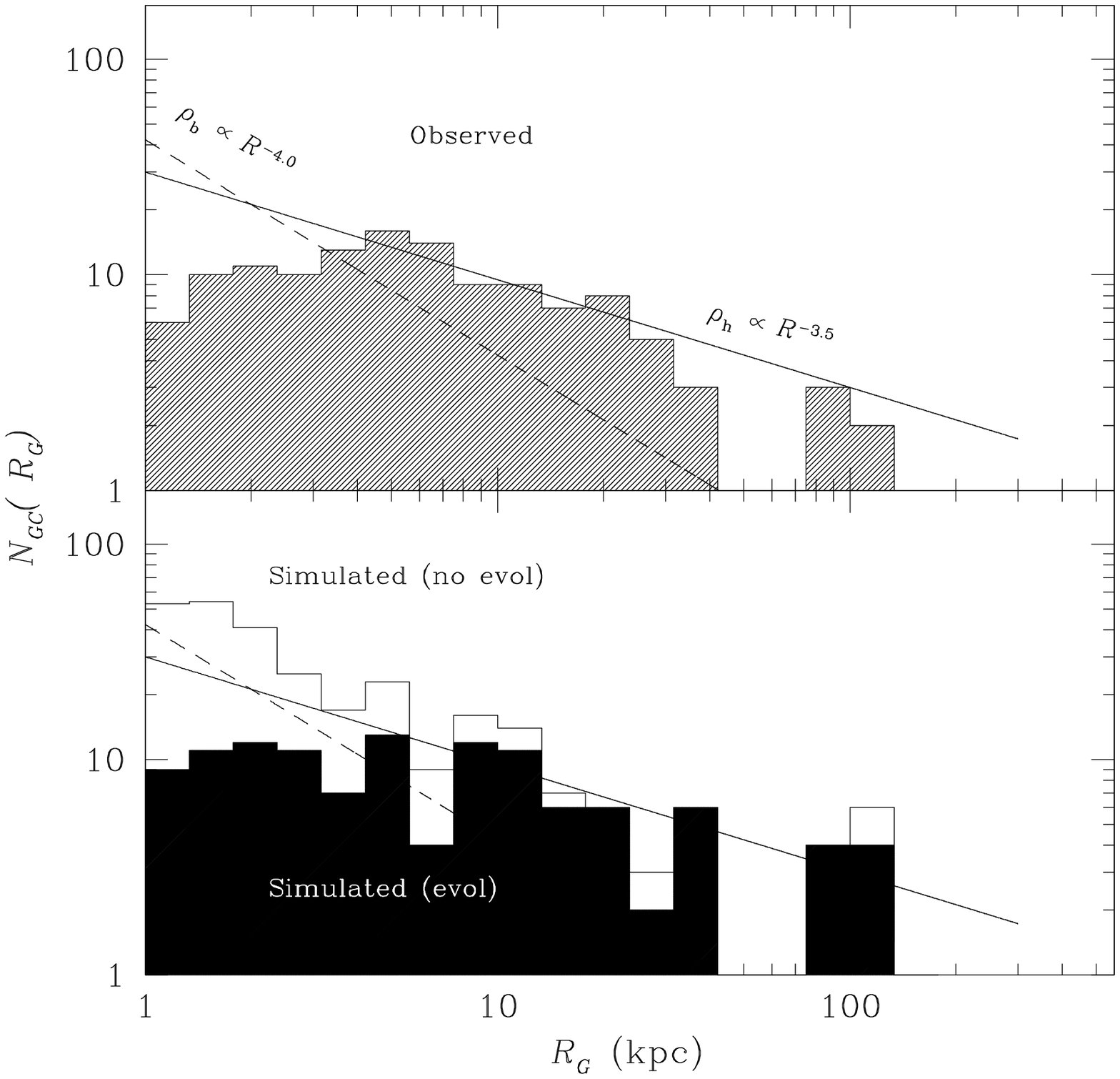}
\figcaption[mws.06.ps]{
(Upper Panel) Distribution of Galactocentric radii for Galactic globular clusters. 
The radial profiles corresponding to the density laws $\rho_h (R_G) \propto R_G^{-3.5}$
and $\rho_b (R_G) \propto R_G^{-4.0}$ are indicated by the solid and dashed lines, 
respectively. The curves have been arbitrarily scaled to match at $R_G = 2$ kpc.
(Lower Panel) Radial distribution of globular clusters for the simulation shown 
in Figure 4. The open histogram shows the initial profile, while the
solid histogram indicates the distribution of the surviving globular clusters. 
\label{fig6}}

\clearpage
\plotone{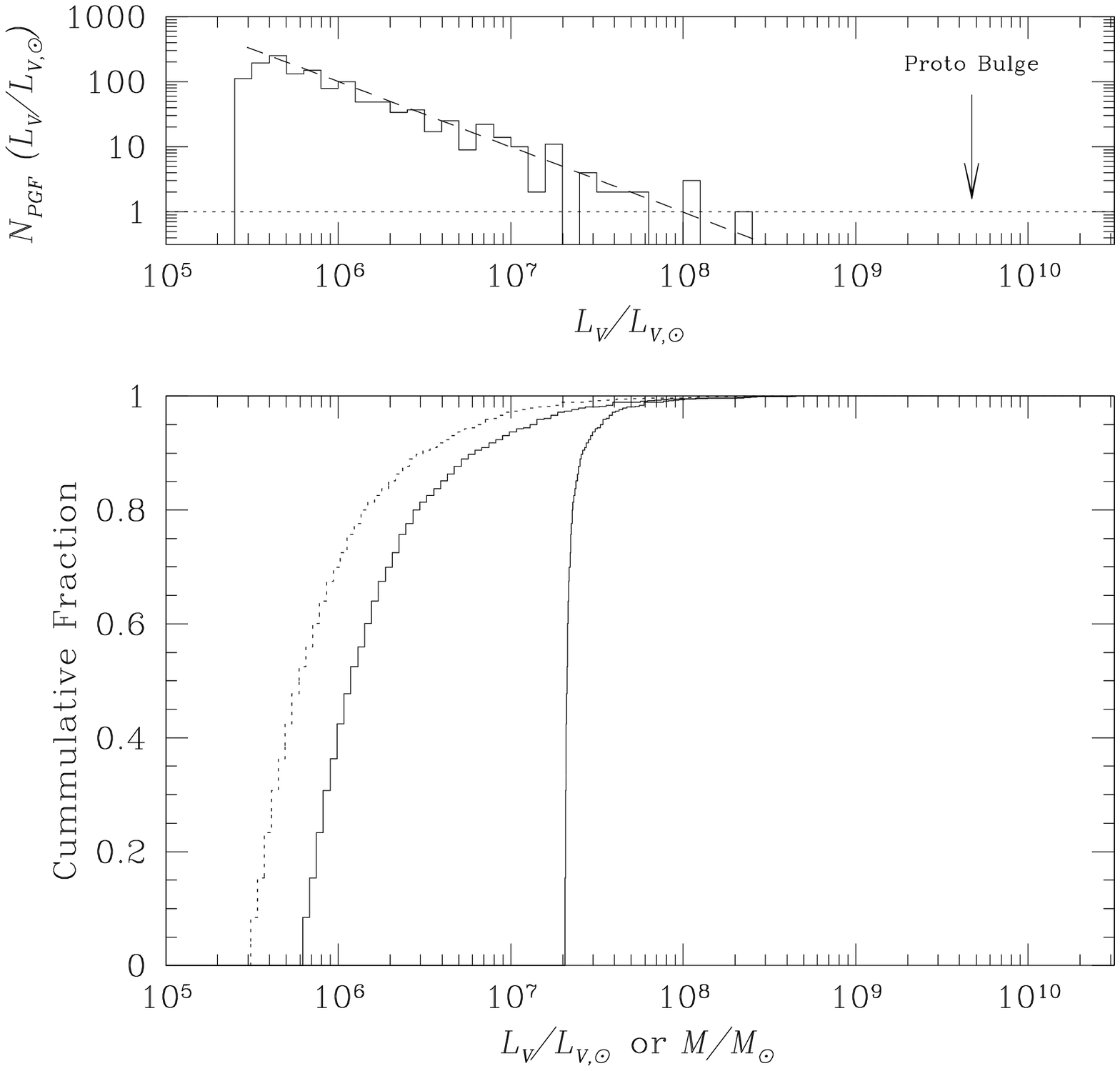}
\figcaption[mws.07.ps]{
(Upper Panel) Luminosity distribution proto-Galactic fragments for the
simulation shown in Figure 4.
A power-law luminosity function with slope $\alpha = -2$ is shown as
the dashed line. The location of the ``proto-bulge" is indicated by the arrow.
(Lower Panel) Cumulative distribution of proto-Galactic fragment luminosities 
for the same simulation (dotted curve). The two solid curves show the
cumulative distribution of proto-Galactic fragment masses assuming: (1) 
$M = 2L_{V}(M_{\odot}/L_{V,{\odot}})$;
and (2) $M = 2L_{V}(M_{\odot}/L_{V,{\odot}}) + 2{\times}10^7 M_{\odot}$.
\label{fig7}}

\clearpage
\plotone{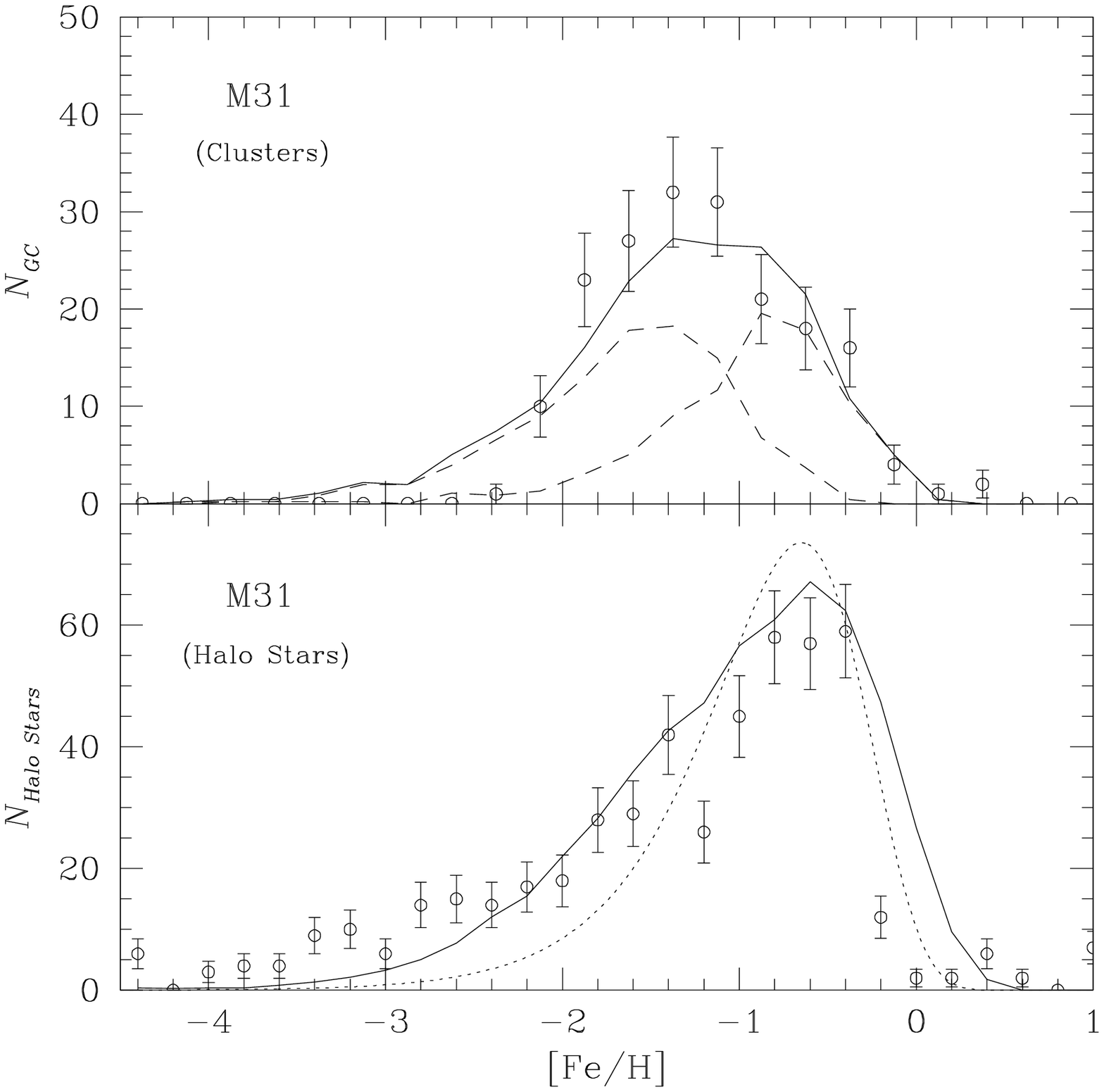}
\figcaption[mws.08.ps]{One simulation of the M31 globular cluster metallicity distribution.
Dynamical effects have not been included. The separate metal-rich and metal-poor components
are indicated by dashed curves; the combined distrubution is indicated by the solid curve.
The open circles show the actual distribution using the cluster sample of 
Huchra, Brodie \& Kent (1991).
(Lower Panel) Corresponding distribution of halo field star metallicities
based on this simulation (solid curve). The halo star metallicities come
from the G312 field of Holland, Richer \& Fahlman (1996).
The dotted line shows the prediction of the Simple Model for an effective yield
of ${\log{y_e}} = -0.65$.
\label{fig8}}

\clearpage

\plotone{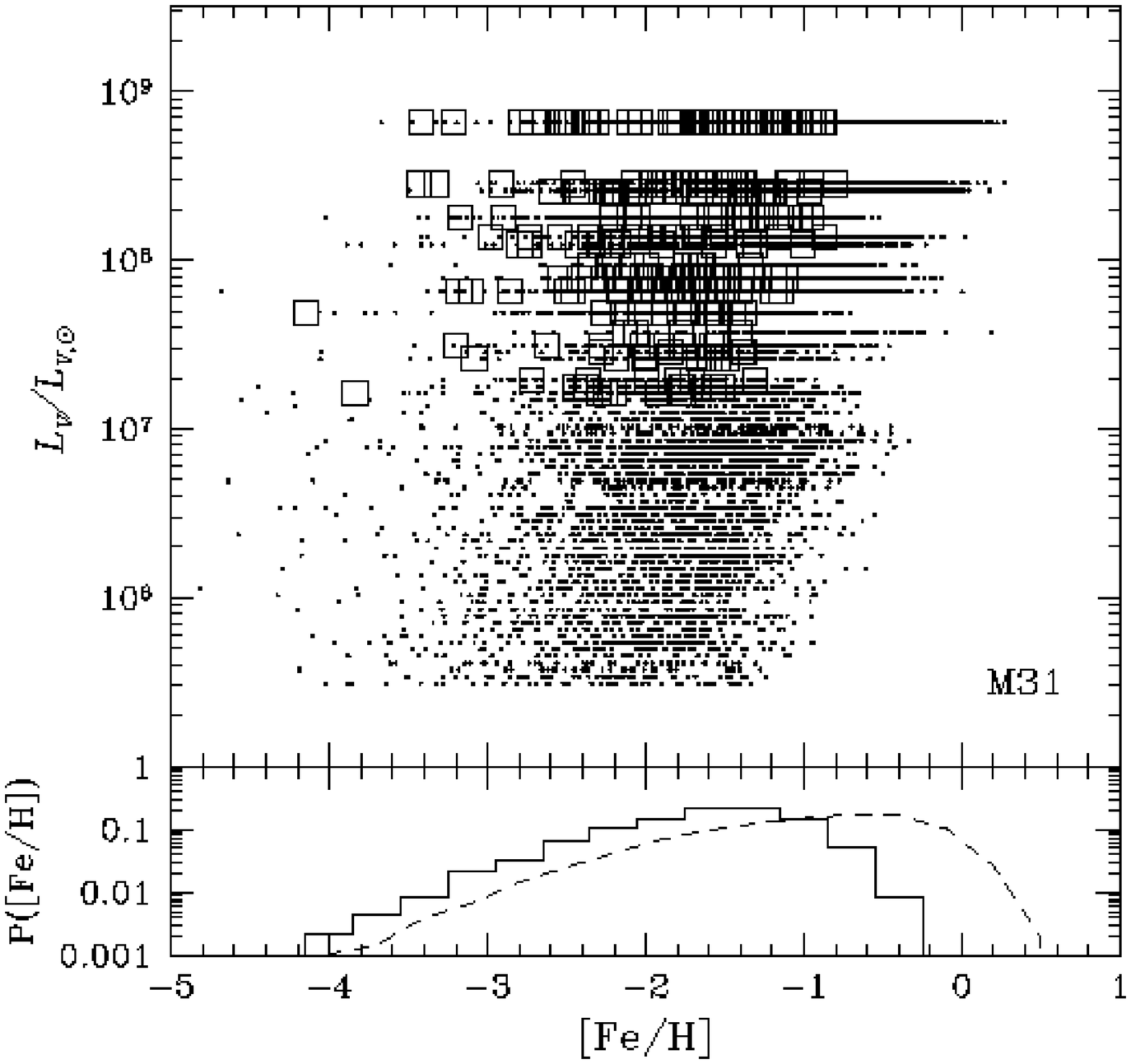}
\figcaption[mws.09.ps]{Same as Figure 5, except for simulation of the M31
spheroid shown in the previous figure. Note that the M31 halo field star
population is clearly more metal-rich than its globular cluster system
due to the smaller number of low-luminosity fragments incorporated into
the spheroid.
\label{fig9}}

% Option 2.  The figure captions are printed on a caption page(s) as in 
% option 1.  The figures available as EPS files are then printed at the
% end of the document, one figure per page, using the \plotone command.
% If you wish to process this option then simply comment out the \end{document}
% just above these five lines. 
 
\end{document}